\documentclass[reprint,amsmath,amssymb,aps,pre]{revtex4-2}
\usepackage{graphicx}
\usepackage{dcolumn}
\usepackage{bm}
\usepackage{url}
\usepackage {hyperref}
\DeclareUnicodeCharacter{0306}{ }

\begin{document}
\preprint{IISER/123}

\title{Biased random walkers and extreme events on the edges of complex networks}
\author{Govind Gandhi and M. S. Santhanam}
\affiliation{Indian Institute of Science Education and Research, Dr. Homi Bhabha Road, Pune, India - 411008}

\date{\today}

\begin{abstract}
Extreme events have low occurrence probabilities and display pronounced deviation from their average behaviour, such as earthquakes or power blackouts. Such extreme events occurring on the nodes of a complex network have been extensively studied earlier through the modelling framework of unbiased random walks. They reveal that the occurrence probability for extreme events on nodes of a network has a strong dependence on the nodal properties.
Apart from these, a recent work has shown the independence of extreme events on edges from those occurring on nodes. Hence, in this work, we propose a more general formalism to study the properties of extreme events arising 
from {\it biased} random walkers on the edges of a network. This formalism is applied to biases based on a variety network centrality measures including PageRank. It is shown that with biased random walkers as the dynamics 
on the network, extreme event probabilities depend on the local properties of the edges. The probabilities are highly variable for some edges of the network, while they are approximately a constant for some other edges on 
the same network. This feature is robust with respect to different biases applied to the random walk algorithm. Further, using results from this formalism, it is shown that a network is far more robust to extreme 
events occurring on edges when compared to those occurring on the nodes.
\end{abstract}

\maketitle

\section{Introduction}
\label{sec:intro}
Extreme events arise in many situations of practical interest \cite{albeverio2006extreme}. The most common ones are the extremes witnessed in geophysical phenomenon such as the earthquakes, cyclones and drought. These represent extreme events recorded in a univariate time series. However, networks have emerged as a standard modeling paradigm to describe such complex systems \cite{vespignani2012modelling,barabasi2012network}. For instance, traffic jams, large scale power blackouts, call drops in cellular phone networks, overloaded web servers are a result of large fluctuations beyond the capacity of the system to service the requests. These events take place, respectively, on a
network of roads, power distribution network, cellular network and on a network of routers.
In general, any large excursion of a dynamical variable from its typical behaviour can be termed 
an extreme event. Though such events typically have low probability
of occurrence, they are regarded as significant due to the social and financial losses suffered on 
account of most of these extreme events.

While the classical extreme value theory \cite{gumbel1958statistics}, formulated a century ago, deals with extreme events primarily 
in univariate time series, the study of extreme events on complex networks has attracted attention 
only since the last decade \cite{chen2013discovery,chen2014controlling,mahecha2017detecting}. In the latter case, a typical setting is that of a topology of a complex network, a system of nodes connected through edges \cite{albert2002statistical}. In order to discuss events and extreme events, a dynamical process needs to be additionally defined on the networks. If we then let $x(t)$ represent the flux of any dynamical variable on a node of a network, then an event is deemed to be an extreme event if $x(t) > x_{\text{th}}$, where $x_{\text{th}}$ is a suitable threshold that indicates, for instance, 
the processing capacity of a node.  An early work along this direction used non-interacting random walks as a dynamical
process \cite{kishore2011extreme}. By assuming $x_{\text{th}}$ to be proportional to the typical random walker flux on a node,
it was demonstrated that the probability that an extreme event occurs on that node depends on the
degree of the node in question. In particular, quite contrary to intuition, it was observed that,
on an average, hubs have low probability for occurrence of extreme events, while small degree nodes
have higher likelihood for extreme events \cite{kishore2011extreme}. Over the years, this proposition was observed in
many different settings. 
For instance, this feature been theoretically reported in a network of coupled 
chaotic maps \cite{Moitra_sinha_2019}, in stochastic processes such as Brownian motion in 
an external potential \cite{Amritkar_Ma_Hu_2018}, and in biased random walks on networks \cite{kishore2012extreme}.
However, it appears that in the case of interactions among agents, the extreme events created
by the agents can possibly have a different profile for the probability for the occurrence of
extreme events \cite{chen2015extreme, gupta2021extreme}. The formulation introduced in Ref. \cite{kishore2011extreme}
has been used to study network failures as well \cite{mizutaka2013structural, mizutaka2014network}.
In general, the focus of these works has been on extreme events, specifically on the nodes of a network.

A natural extension of these works is to consider extreme events occurring at the interface where nodes interact - the edges of the network. Recently, using random walks as flows on a network, extreme events
occurring on edges and their properties were studied \cite{kumar2020extreme}.
Let $i$ and $j$ be two nodes connected by an edge $e_{ij}$. It was shown that 
extreme events on nodes $i$ and $j$, and those on the edge $e_{ij}$ connecting the nodes 
are nearly uncorrelated and that they should be regarded as events which do not influence one another \cite{kumar2020extreme}. 
Hence results for the occurrence of extreme events on nodes do not characterise those taking place on the 
edges. Indeed, most of our personal experience in the case of road traffic tells us that traffic jams on 
junctions and at any other point on the roads are not always correlated.  Furthermore, with unbiased walkers occurrence probability of extreme events on edges depends only on the gross network parameters such as
total number of edges and not on the local details in the vicinity of the edge. This feature differs appreciably from what we know about extreme events on nodes. Motivated by these considerations we study extreme events on the edges using the more broad dynamics of {\it biased} random walkers on complex networks. The standard random walk provides
a simple and yet non-trivial setting to explore this question. In the rest of this paper, a general formalism
is developed to study biased random walks on networks, and it is applied to the degree biased 
random walks on scale-free networks, and also to other centrality measure based biases such as 
eigenvector centrality, betweenness  centrality and page-rank centrality.  It is also applied to a 
real-life planar street network. In all the cases studied here, extreme events arise from inherent fluctuations in the number of walkers, and provides insights into how inherent fluctuations, as opposed to external shocks, give rise to extreme events.

 We describe the formalism in sections \ref{BRW_intro}, \ref{BRW_nodes}, \ref{BRW_edges}. In section \ref{DBRW} we apply it to a planar network and subsequently to several different types of biased random walkers. Finally, in section \ref{stability} we discuss the robustness of edges against occurrence of extreme events.

\section{Biased Random Walks on networks} \label{BRW_intro}

A connected, undirected network with $N$ nodes and $E$ edges is considered. Its connectivity
structure is described by an adjacency matrix ${\mathbf A}$ whose elements 
$a_{ij}$ are either $1$ or $0$ depending on whether an edge connects nodes $i$ and $j$ or not. 
On this network structure, diffusion of $W$ independent random
walkers is considered whose dynamics is biased by any network property that can be associated with the node, such as the degree. This is represented by a single-step transition probability $\pi_{ij}$ for a walker to go from node $i$ to node $j$, while satisfying the normalization condition $\sum_j \pi_{ij}=1$ for all $i$. In this work, we study a
class of biased random walks defined by the one-step transition probability
\begin{equation}
\pi_{ij}=\frac{a_{ij}f_{j}}{\underset{m}{\sum}a_{im}f_{m}}, \;\;\;\;i,j=1,2, \dots, N.
\label{transition prob matrix}.
\end{equation}
The biasing function $f_{j}$ here, could represent any structural or non-structural property of the network as long as it is time-independent and is associated with a node. The structural properties could simply be the degree of a node, the betweenness centrality, it could be any amalgam of structural features of the node and its local neighbourhood, it could be the ground-truth community the node is externally assigned, and so much more. This can be generalized to the case of multiplex 
networks as well \cite{battiston2016efficient}. In the case of unbiased random walk $f_{j}=1$,  for all $j$, indicating that the probability for a walker to jump to any of its neighbouring nodes is a constant.

Let $P_{ij}(t)$ denote the probability that a walker at node $i$ starting at time $t=0$ 
reaches node $j$ after $t$ discrete jumps. The time evolution of probability is 
given by the master equation
\begin{equation}
P_{ij}(t)=\sum_{k=1}^{N} P_{ik}(t-1) ~ \pi_{kj}.
\end{equation}
By allowing only time independent biasing functions we are assured transition probabilities $\pi_{ij}$ that are stationary and that any node in the network is accessible from any other node via a finite number of steps. It then follows from Perron-Frobenius theorem \cite{ninio1976simple} that for any such biasing function $f_{j}$, the stationary probability distribution $p_{i}$ exists. This time-independent occupation probability on node $i$ turns out to be \cite{noh2004random}:
\begin{equation}
p_{i}=\frac{f_{i} \sum_j a_{ij} f_{j} }{\underset{k}{\sum} f_{k} \left( \underset{j}{\sum} a_{kj}f_{j} \right) },
\label{occprob1}
\end{equation}
in which the normalization condition $\sum_{j}p_{j}=1$ holds. Note that the occupation probability $p_i$
does not restrict the form of the biasing function or the network structure at this stage, thereby making it sufficiently general.

\section{Extreme Events on nodes} \label{BRW_nodes}
The occupation probability in Eq. \ref{occprob1} can be used to
evaluate extreme event probabilities on nodes. The probability of 
finding $w>1$ non-interacting walkers on the node labelled $i$ is $p_{i}^w$ while the rest of the $W-w$ walkers are distributed among the other nodes of the network. The distribution of number of walkers on node $i$ is then a binomial distribution given by \cite{kishore2011extreme}
\begin{equation}
\mathcal{P}_i(w)=\binom{W}{w} ~ p_{i}^w ~ \left(1-p_{i} \right)^{W-w}.
\label{TraffDistNode}
\end{equation}
Intuitively, an extreme event represents a pronounced deviation from the mean flux.  
The flux $w$ represents events in our framework and it is deemed to be an extreme 
event if $w \geq q_i$, where $q_i$ represents a suitable threshold.
Then, using Eq. \ref{TraffDistNode}, the EE probability on $i$-th node can be
defined as
\begin{equation}
\mathcal{P}_{EE}(q_{i}) = \sum_{w=q_{i}}^{W} \mathcal{P}_i(w)\label{PEEN}.
\end{equation}
The threshold $q_{i}$ is chosen to be proportional to the natural variability 
of the flux passing through the node. As was done earlier in Ref. \cite{kishore2012extreme},  in this work 
too the threshold for identifying EE is taken to be
\begin{equation}
q_{i}=\langle w_{i}\rangle + m \sigma_{i},
\label{PEEN threshold}
\end{equation}
where $m \geq 0$. The average and the variance of the flux passing through $i$-th node are given as
\begin{subequations}
\begin{align}
\langle w_{i}\rangle & =Wp_{i}, \label{TraffMean} \\
\sigma_{i}^{2} & = W p_{i} \left(1-p_{i} \right). \label{TraffVar}
\end{align}
\end{subequations}
Since these quantities depend only on the biasing function and the local structural information about immediate neighbours of nodes, these results will hold good 
for any network irrespective of its degree distribution.
This notion of EE is sufficiently general to encompass systems that accommodate congestion, {\it e.g.}, airplane/vehicular traffic in a network of airports/streets.

\section{Extreme Events on Edges} \label{BRW_edges} 
\label{EEE}
While extreme events on nodes have been studied earlier, there has not been much work about EE on the edges of a network. As was recently shown in Ref. \cite{kumar2020extreme}, the occurrence of EE on nodes and
its edges are nearly uncorrelated. Hence, it is not possible to infer about EE on edges/nodes based on the
information about EE on nodes/edges. Physically, this is intuitive because once an EE takes place on a node, the walkers get scattered into a large number of edges and might not necessarily lead to an EE on the edges. These events on the edges will depend on the flux through edges and the carrying capacity
of the edges. For instance, in the water supply network, the reservoirs might have ample capacity to hold the water, though its transport channels might have a smaller \textit{load}-bearing capacity. A microprocessor has a network of logic gates, where the information is carried as electrical signals. In all these cases, the conduits (edges) can only support a finite range of fluctuations \cite{barabasi2004fluctuations}, beyond which it should be designated as an extreme event. In most cases, the occurrence of EE on the edges can lead to disasters. Therefore it is imperative that EE on edges be studied on its own right.

\subsection{Load and Flux Distribution on edges}
Firstly, we seek quantifiable definitions of the situations described above, {\it i.e.}, for load and flux. 
Consider an edge $e_{ij}$ and the nodes it connects labelled $i$ and $j$. Suppose there are $w_{i}$ walkers 
on node $i$ and $w_{j}$ walkers at node $j$. At the $t$-th time step, let $l_{i}$ out of the $w_{i}$ walkers jump into node $j$, and $l_{j}$ out of the $w_{j}$ walkers jump to node $j$. The load $l_{ij}$ and flux $f_{ij}$ of 
the edge $e_{ij}$ at time $t$ is then defined as
\begin{equation}
l_{ij}(t)=l_{i}(t)+l_{j}(t)  \,\,\,\,\,\,\,\,\,\,\,\,\, f_{ij}(t)=l_{i}(t)-l_{j}(t)\label{fload}
\end{equation}
Since the load $l_{ij}$ is stochastic, a natural question of interest is its
probability distribution for a particular edge $e_{ij}$. The details of obtaining 
this probability distribution of load for a given edge $e_{ij}$ is given in the 
appendix \ref{AppB}. The normalized probability distribution comes out to be
\begin{equation}
P_L(l_{ij})=\binom{W}{l_{ij}} ~ (2p_{i} \pi_{ij})^{l_{ij}} ~ (1-2p_{i} \pi_{ij})^{W-l_{ij}}.
\label{prdistload}
\end{equation}
Comparing this with Eq. \ref{TraffDistNode}, an effective occupation probability on edge $e_{ij}$ 
can be identified as $p_{ij}^{\text{eff}} = 2 p_{i} \pi_{ij}$. Physically, this corresponds to the product of the probability of a walker to be on node $i$  and probability for that walker to move into node $j$. Therefore, the distribution of load on an edge is dependent on the
product of the biasing functions of the two nodes involved in the
edge. It is dependent on the local network structure and not on the global network parameters. 
The mean load and the variance, respectively, on the edge $e_{ij}$ are
\begin{subequations}
\begin{align}
\langle l_{ij}\rangle & = 2Wp_{i} \pi_{ij} = W p_{ij}^{\text{eff}} = \frac{2Wa_{ij}f_{i}f_{j}}{{\sum_{l}}({\sum_{k}}(a_{lk}f_{k})f_{l}},
\label{loadmean}\\
\sigma_{ij}^{2} & = W p_{ij}^{\text{eff}} ( 1 - p_{ij}^{\text{eff}} )
\label{loadstdev}
\end{align}
\end{subequations}
In the limit that $ p_{ij}^{\text{eff}} \ll 1$, then it is easy to see that 
$\sigma_{ij}^2 \approx \sqrt{\langle l_{ij} \rangle}$, a result known in the case
of unbiased random walks on networks.

For the probability that a node $i$ has $l_i$ walkers, we make a similar calculation done in Appendix \ref{AppB} to get
\begin{equation}
\mathcal{L}(l_{i};\,i)=\binom{W}{l_{i}} ~ (p_{i} \pi_{ij})^{l_{i}} ~ (1-p_{i} \pi_{ij})^{W-l_{i}}
\label{eq:li}
\end{equation}
The flux distribution {\it i.e.}, the probability that the flux equals $f_{ij}$ on an edge $e_{ij}$ is then product of probabilities of getting different values $l_i$ and $l_j$ constrained by Eq. \ref{fload}. We sum over all such possibilities to get the flux distribution on an edge to be
\begin{equation}
P_F(f_{ij})=\sum_{m=0}^{W-f_{ij}} \mathcal{L}(l_{i}=m+f_{ij};\,i) ~ \mathcal{L}(l_{j}=m;\,j).
\label{flux_dist}
\end{equation}
Using the load (Eq. \ref{prdistload}) and flux distributions (Eq. \ref{flux_dist}), we can now compute the extreme events on edges for a variety of networks.

\subsection{Probability of EE on edges for load}
Similar to the definition of EE on nodes, if the load on an edge at any time $t$ exceeds a defined threshold $q_{ij}$ such that $l_{ij}(t) \geq q_{ij}$, then it indicates the occurrence of load EE on edge $e_{ij}$. 
As in the case of nodal extreme events, $q_{ij}$ is taken to be 
proportional to the natural variability of the load on the edge. Thus,
\begin{equation}
q_{ij}=\langle l_{ij}\rangle+m\sigma_{ij}\label{thresh-1},
\end{equation}
where $m\geq0$, and  the mean load $\langle l_{ij}\rangle$ and standard
deviation $\sigma_{ij}$ are given by Eqs. \ref{loadmean} and \ref{loadstdev} respectively which inform us that $q_{ij}$ is a function of  $p_{ij}^{\text{eff}}$. Hence, the probability for load EE on an edge $e_{ij}$ also depends on $p_{ij}^{\text{eff}}$. Then by using Eq. \ref{prdistload} and Eq. \ref{loadmean} we obtain for probability of 
load EE on an edge $e_{ij}$
\begin{equation}
Q_L \left( \langle l_{ij} \rangle \right) = \sum_{l=q_{ij}}^{W} P_L(l).
\label{eenode-1}
\end{equation}
The probability for the occurrence of flux EE, $Q_F \left( \langle f_{ij} \rangle \right)$, 
on an edge is quite similar to Eq. \ref{eenode-1}, with $P_L(l)$ replaced by flux 
distribution $P_F(f)$ from Eq. \ref{flux_dist}. In general, the EE probabilities on
edge $e_{ij}$ depend only on the mean load or flux, which in turn carries information 
about local network structure and strength of bias in random walk dynamics.
Furthermore, $\langle l_{ij}\rangle$ is characteristic of an edge irrespective of the 
network topology, and allows us to compare EE probabilities for different biases 
and even between networks on a common footing. This accessible quantity is a candidate for an observable when studying real life systems.

We are currently in a position to take our pick of biased random walk dynamics and a network structure and obtain information on the EE probabilities on edges and nodes for congestion-type systems. In the rest of the paper, this framework is tested and compared against simulation results for different random walk dynamics and network topologies.

\section{Biased Random Walks} 

\label{DBRW}

\begin{figure}
\includegraphics{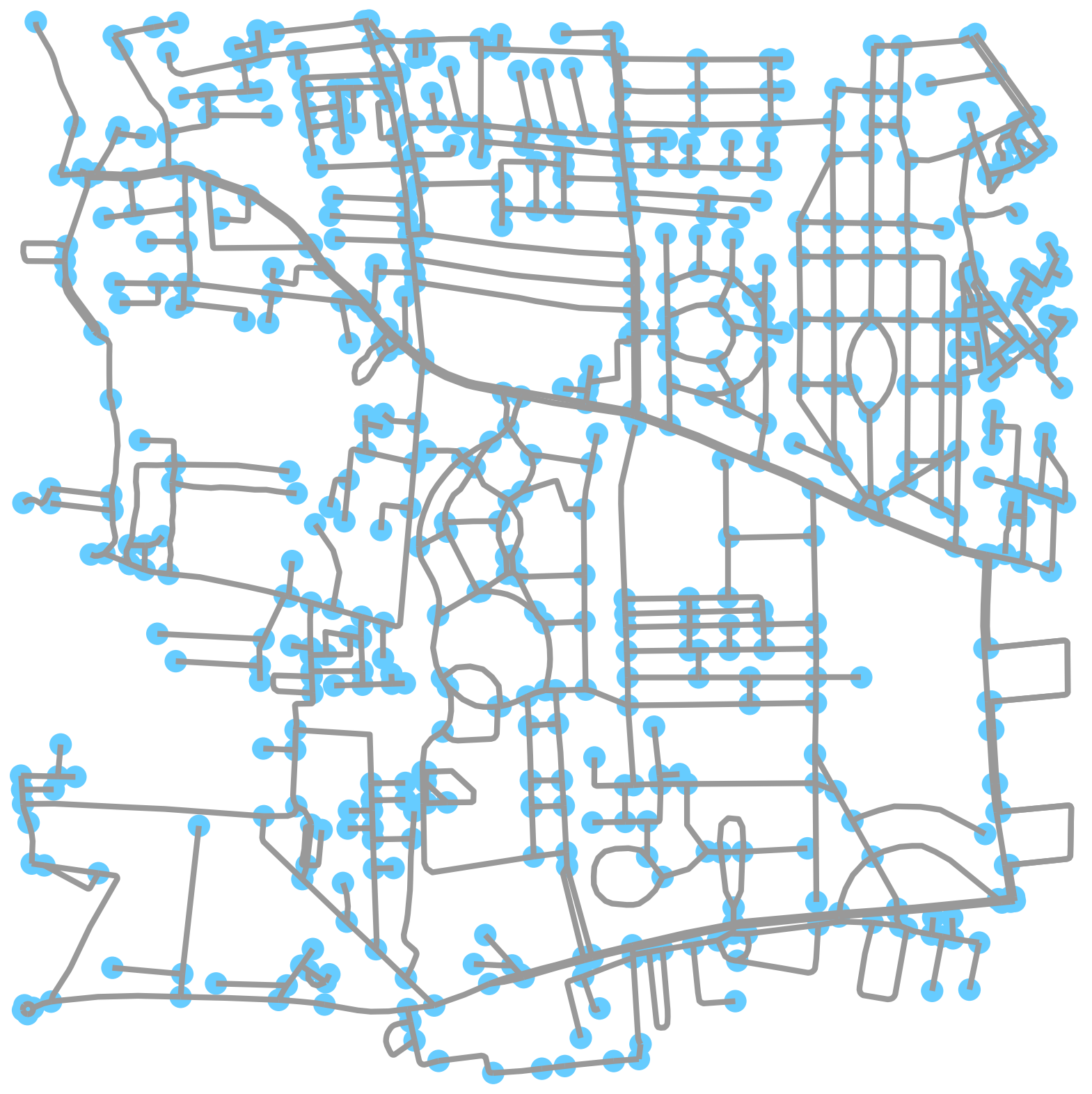}
\caption{The 1.5 km\protect\textsuperscript{2} planar street network around IISER Pune
with 1010 nodes and 1306 edges. The nodes in the two-dimensional plane denote street intersections and the edges represent streets.}
\label{IISERP}
\end{figure}
\subsection{Degree Bias} 
In this section, we shall consider one of the most applicable biases - the degree biased random walk \cite{fronczak2009biased, wang2006traffic, zlatic2010topologically}. Its biasing function is of the form
\begin{equation}
f_{i}=k_{i}^{\alpha},   
\label{eq:degbias} 
\end{equation}
where $k_i=\sum_j a_{ij}$ is the degree of $i$-th node. The exponent
$\alpha$ is the parameter that determines the strength of bias imparted to the walkers.
Unbiased random walk corresponds to $\alpha=0$, whereas for $\alpha > 0$ the walkers
are biased towards hubs (nodes with high degrees), and for $\alpha < 0$ they are preferentially biased towards small degree nodes. $|\alpha|$ determines the strength of the bias towards or away from the hubs. On networks, biased random walks of this type had been studied earlier to enumerate the characteristic time scales, namely, the recurrence time
to the starting point, cover time, and time taken to visit a node for the first time \cite{goldhirsch1987biased,lee2014estimating}. The
parameter $\alpha$ can be tuned so as to minimize these time scales. In general, this results
in smaller time scales with biased random walks in comparison to the unbiased random walks \cite{bonaventura2014characteristic}. The local navigation rules for the biased walks are determined by the transition probability (see Eq. \ref{transition prob matrix}) 
\begin{equation}
\pi_{ij}=\frac{a_{ij} ~ k_{j}^{\alpha}}{\sum_{m}a_{im} ~ k_{m}^{\alpha}}.
\label{eq:dbrw trans}
\end{equation}
The stationary occupation probability follows from Eq. \ref{occprob1}, and is
\begin{equation}
p_{i}=\frac{k_{i}^{\alpha} ~ \underset{m}{\sum}a_{im}k_{m}^{\alpha}}{\underset{l}{\sum} ~ [k_{l}^{\alpha}(\underset{m}{\sum}a_{lm}k_{m}^{\alpha})]}.
\label{DBRW pstat}
\end{equation}
The index in each of the summations runs over all nodes. 

\begin{figure}
\includegraphics{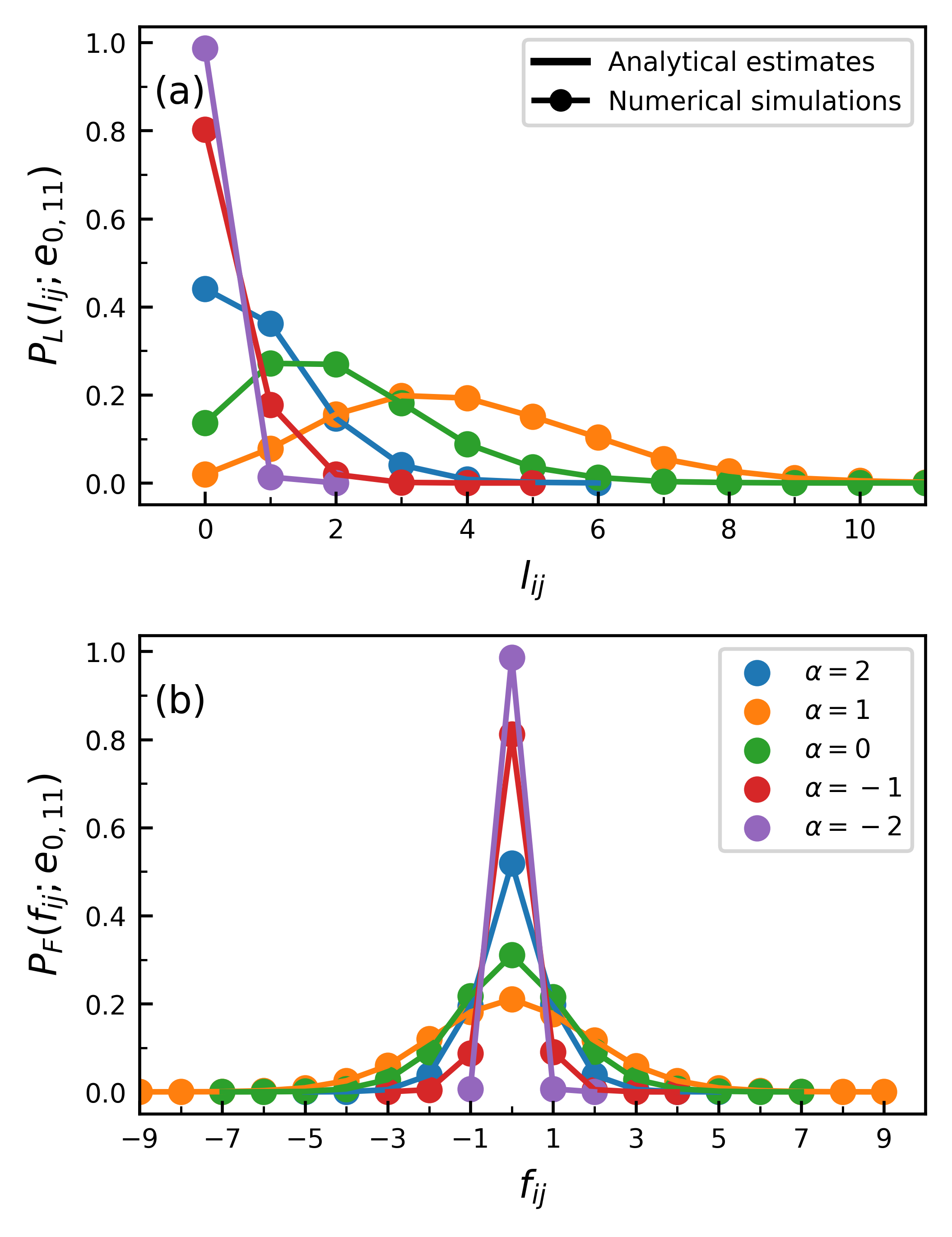}
\caption{The probability distributions of (a) load and (b) flux for a randomly chosen edge $e_{0,11}$ connecting
nodes 0 and 11 with degrees 100 and 44, respectively. The different curves indicate the distributions obtained by executing DBRW for varying values of bias strength $\alpha\in\{-2,-1,0,1,2\}$. Each execution had 2500 walkers running for 100,000 time steps on the network in Fig. \ref{IISERP} and was repeated 15 times with varying initial positions of random walkers.}
\label{F+L}
\end{figure}

The results are first obtained for a (planar) street network \cite{boeing2017osmnx} shown in Fig. \ref{IISERP}. The planarity of the network was confirmed using Kuratowski's theorem \cite{kuratowski1930probleme}.
Fig. \ref{F+L} displays the probabiity distribution for load and flux obtained from
biased random walk simulations for an arbitrary edge $e_{0,11}$ connecting nodes labelled 0 and 11. 
For the purposes of this simulation, $W=2500$ non-interacting walkers executed a 
degree biased random walk (DBRW) on the network in Fig.\ref{IISERP} for 100,000 time steps.
The results shown in Fig. \ref{F+L} represent an average over 15 realisations with a fixed
network, but with varying initial positions of the walker for each realisation.
Evidently, the simulation results are in good agreement with the analytical results in Eqs. \ref{eq:li} and \ref{flux_dist}. These results differ substantially from the corresponding distributions for
unbiased random walks, in which case {\it every} edge on the network has the identical probability 
distribution for load and flux independent of network topology \cite{kumar2020extreme}. 
In the present case of biased random walks, the distribution of flux or load depends on nodal
properties. Futher, as the load distributions reveal, for $\alpha >0$ since walkers preferentially 
move towards hubs, edges with smaller mean loads do not attract walkers. On the other hand, if
strongly biased towards small degree nodes, it is possible for walkers to avoid edges which
connect fairly high degree nodes. Hence, the higher probability of null mean loads as seen in Fig. \ref{F+L} for $\alpha <0$. Thus, by
tuning the bias, an edge can be made to either attract or repel walkers.
The results for $\alpha=0$ (unbiased random walk) agree exactly with the results of Ref. \cite{kumar2020extreme}. However, unlike the case of unbiased random walk where every edge has the same probability distribution, edges with different mean loads have distributions according to Eq. \ref{prdistload}.


\begin{figure}
\includegraphics{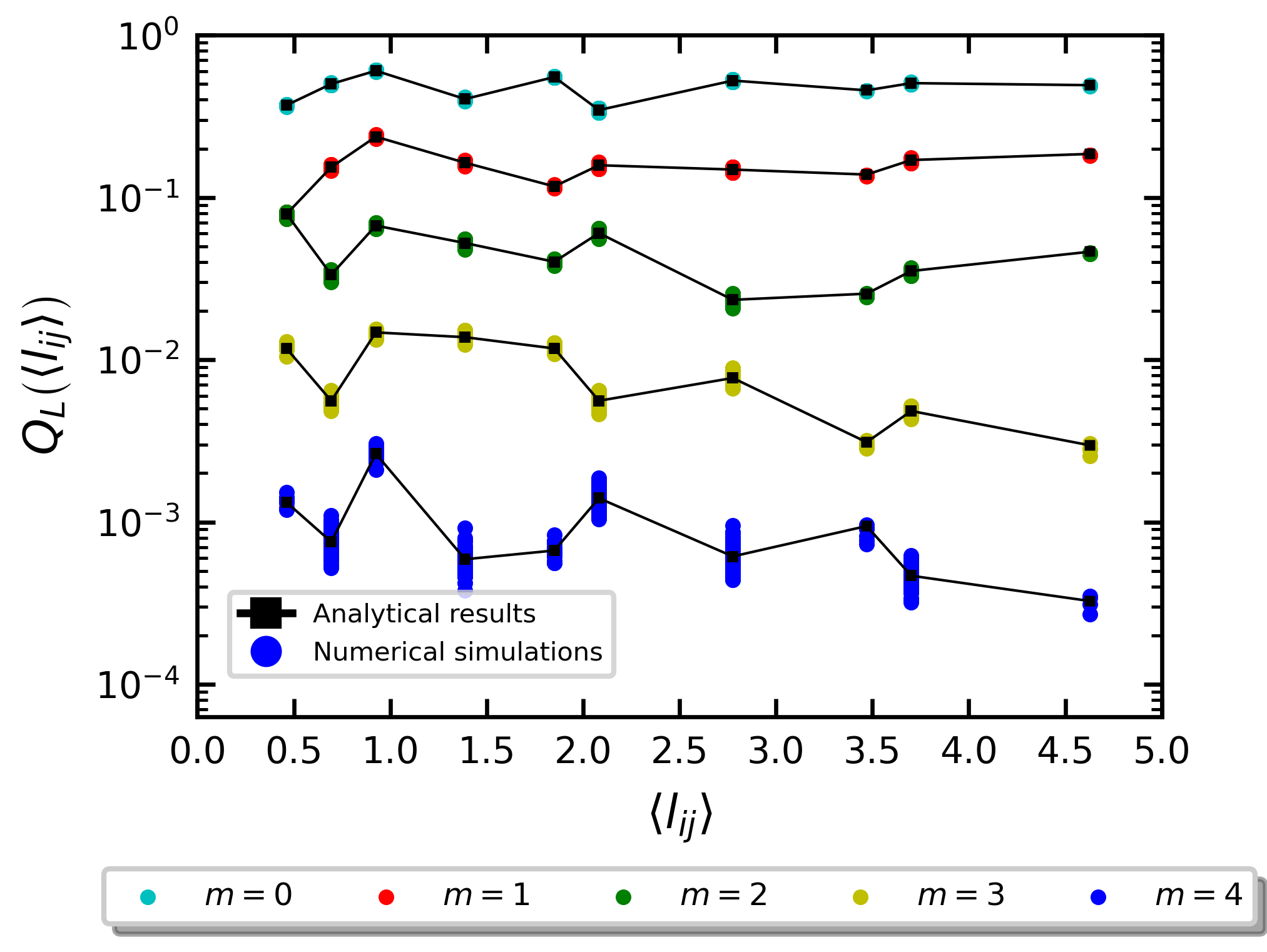}
\caption{Probability of load EE vs mean load on edge for different thresholds of extremeness: $m\epsilon[1,5]$. 2500 walkers performed a DBRW with $\alpha=1$ for 100,000 time steps on the network mentioned in Fig.\ref{IISERP} }
\label{PEEL}
\end{figure}

Next, the probability for the occurrence of load EE, $Q_L \left( \langle l_{ij} \rangle \right)$
on edge $e_{ij}$ is considered. The theoretical result is obtained from
Eq. \ref{eenode-1} and is juxtaposed with simulation results for 
$Q_L \left( \langle l_{ij} \rangle \right)$ in Fig. \ref{PEEL}. It is displayed for all the edges for several threshold values indexed by $m$ at a fixed value of bias ($\alpha=1$). An excellent agreement is observed between the analytical results and the simulations.
In an average sense, the EE probability is higher for edges with lower values of 
$\langle l_{ij} \rangle$ in comparison to edges with larger $\langle l_{ij} \rangle$.
This effect is more pronounced for higher EE thresholds. As $m \to 0$,
the mean load itself becomes the threshold value effectively leading to most events designated
as extreme. In this case, EE probability does not vary much with $\langle l_{ij} \rangle$. Clearly, this quantity varies with $\langle l_{ij} \rangle$ but does not vary much as $m$ is varied. Hence, for a fixed value of threshold indexed
by $m$, EE probability has been averaged over different number of extreme events. It was also found that the density of edges for each value of $\langle l_{ij}\rangle$ differed, but for a given value of $\langle l_{ij}\rangle$, the relative density remained the same for all values of $m$. This is not a peculiarity of biased random walks but is reflective of the inhomogeneity of the network structure.

Now that we have an understanding of the effect of the choice of threshold we proceed to analyse how changing the bias strength affects the system. Fig. \ref{PEEL_2sigma} shows $Q_L \left( \langle l_{ij} \rangle \right)$ plotted against the mean load  $\langle l_{ij}\rangle$ for varying values of $\alpha$
at fixed EE threshold of $m=2$. We have run it on a scale-free network with 2000 nodes and 7984 edges. On this network, 15968 non-interacting walkers execute degree biased random walk for 75000 time steps (note that these are the parameters and network used for the rest of the graphs, unless mentioned otherwise). An excellent agreement between analytical and numerical simulations can be seen. Note that for the 
unbiased random walk dynamics ($\alpha=0$) shown in Fig. \ref{PEEL_2sigma}(c) all the edges have the 
same mean load and hence identical EE occurrence probability, in agreement with the earlier
result \cite{kumar2020extreme}. The effects of biasing is visible in the figure. When biased towards
hubs with $\alpha >0$, more walkers moving towards the hubs leads to EE on edges which connect two hubs. At the other end of the spectrum, when $\alpha < 0$, walkers preferentially move towards small degree nodes and consequently fewer or no EE takes place on edges connecting hubs. The discontinuity in $Q_L \left( \langle l_{ij} \rangle \right)$ arises due to the discreteness of walkers while the threshold $q_{ij}$ in Eq. \ref{eenode-1} is a real number.

In general, considerable variability in the EE probability can be observed. As Fig. \ref{PEEL_2sigma}(e)
shows for $\alpha =-2$, for mean load $\langle l \rangle < 1$, 
EE probability varies by 
4 orders in magnitude. For mean loads of $\langle l \rangle > 1$ (yellow background), the variability 
is about one order of magnitude. Similar wide range of variabilities are observed for other values of 
biases shown in Fig. \ref{PEEL_2sigma}. The exception is the case of $\alpha=0$ corresponding to 
{\it unbiased} random walk, for which the mean load is same on all the edges of the network and 
consequently the EE probability is identical on all the edges \cite{kumar2020extreme}. The variability
in EE probability becomes more pronounced for $\alpha < 0$. This is a new feature not encountered in the
EE probabilities on nodes or edges of networks on which unbiased random 
walks are executed \cite{kishore2011extreme,kishore2012extreme}. It would imply the following: 
on the edges characterised by {\it large} mean loads, the extreme events are approximately equiprobable irrespective of its local network structure. On edges with {\it small} mean loads, this probability varies by large orders of magnitude. Hence, two edges whose mean loads do not appreciably differ from one another can have very different EE probabilities. So as far as EE are concerned, biased walker dynamics on edges leads to the network displaying an approximately logical (not necessarily physical) phase separation in to two parts -- one in which EE have only small variations about an average value, and the other in which
large excursions about the mean are observed. This will have implications for strategies that try to harness extreme events. Finally, it must be emphasised that this variability is not due to insufficient averaging but an inherent feature of biased dynamics in networked system.

\subsection{Other Biases}

\begin{figure}
\includegraphics{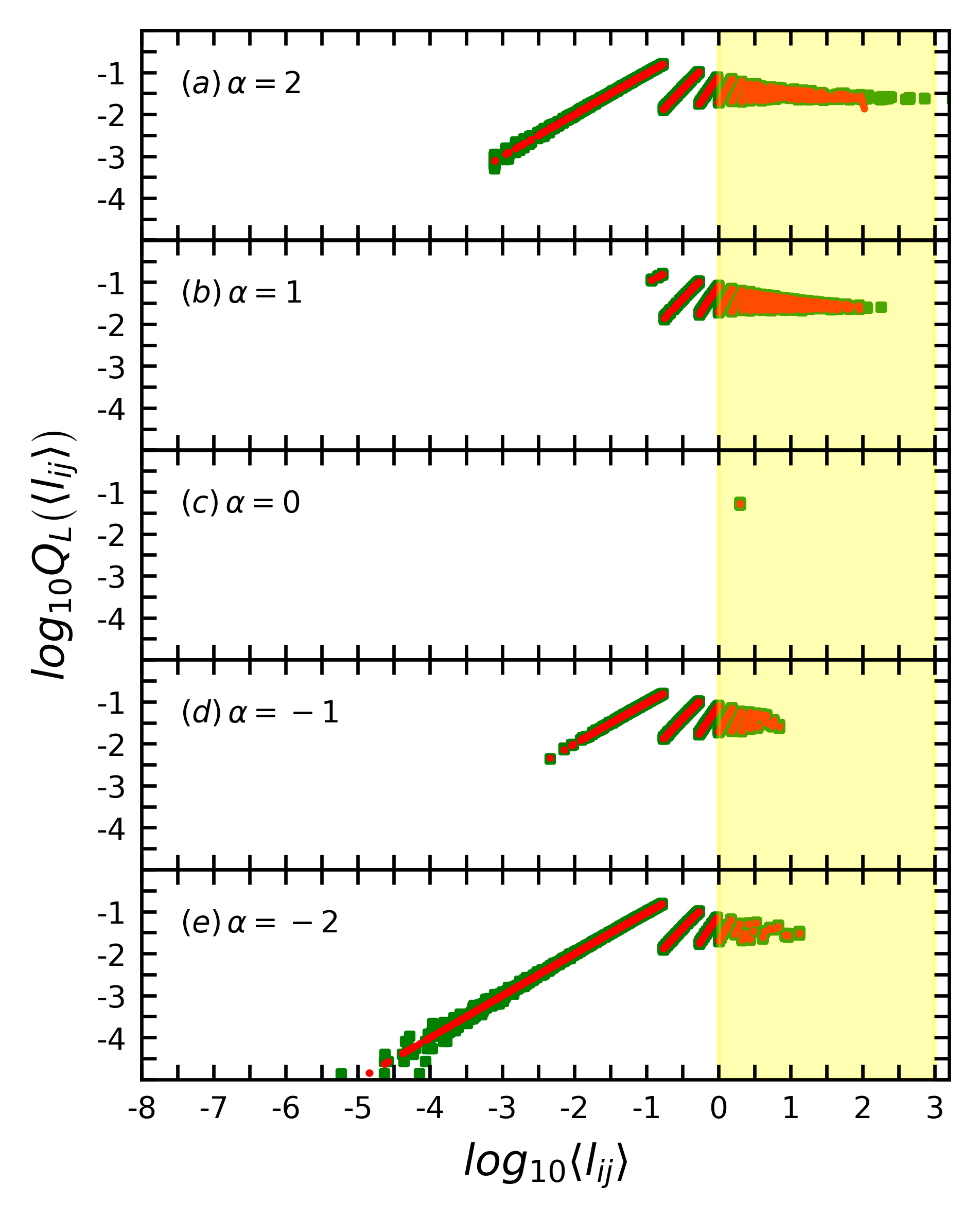}
\caption{Log-log plot of the probabilities of edge EE for load vs mean edge load for different values of bias parameters (a) $\alpha=2$, (b) $\alpha=1$, (c) $\alpha=0$, (d) $\alpha=-1$, (e) $\alpha=-2$ and (f) $\alpha=-4$. The red squares represent analytical results and the green circles indicate numerical simulations. All runs have $m=2$, and were carried out for 75,000 time steps each. The network used was a scale-free network with 2000 nodes and 7984 edges with 15968 non-interacting biased random walkers.}
\label{PEEL_2sigma}
\end{figure}

In this subsection, we switch to a scale-free network \cite{barabasi1999emergence} and apply different kinds of biased walks based on network centrality measures. A scale-free network with 2000 nodes and 7984 edges is created. On this network, 15968 non-interacting walkers executed random walk dynamics for $75000$ time steps. The walkers are randomly distributed on the nodes of the network initially. Simulation data is recorded after discarding the first $10^3$ time steps on account of transient behaviour. 

Since centrality measures provide different ways to rank the importance of nodes in a network, they can also be used as biasing functions. We chose four measures - Betweenness, Eigenvector, Closeness and PageRank centralities \cite{brandes2001faster,bonacich1987power,freeman1978centrality,page1999pagerank} and repeated the procedure followed in section \ref{DBRW}. The Betweenness centrality $B_v$ of a node labelled $v$ is given by
\begin{equation}
B_v = \sum_{a,b} \frac{\sigma_{a,b}(v)}{\sigma_{a,b}},
\end{equation}
where $\sigma_{a,b}$ is the total number of shortest paths from node $a$ to $b$, while $\sigma_{a,b}(v)$ is the number of such paths passing through node $v$. The summation is over all pairs of nodes $(a,b)$ on the network such that $a \ne b \ne v$. It represents the degree to which nodes stand between each other. The eigenvector centrality defines the importance of a node in terms of the neighbourhood of each node.
Let $\mathbf{A}$ be the adjacency matrix, then the Eigenvector centrality
is given by the left eigenvector $\mathbf{x}$ corresponding to the largest eigenvalue $\lambda$:

\begin{equation}
\lambda \mathbf{x} =  \mathbf{x} \, \mathbf{A}.
\end{equation}

The $v$-th element of $\mathbf{x}$ (denoted by $X_v$) gives the centrality measure for the node $v$ on a network. The closeness centrality is the average length of the shortest path between the node and all other nodes in the graph. If $d(v,u)$ represents the length of shortest path between nodes
$v$ and $u$, then closeness centrality is defined as
\begin{equation}
C_v = \frac{N}{\sum_u d(v,u)}.
\end{equation}
Thus, if $C(v)$ is large, then it implies that most $d(v,u)$ are small, and hence node $v$ is "closer" to other nodes on the network. For the present purposes, the biasing function for the $v$-th node is taken to be 
\begin{equation}
f_v \propto \Phi_v^{\alpha},
\end{equation}
where $\Phi_v=B_v,\, X_v$, $C_v$ or PageRank centrality \cite{page1999pagerank}. The transition probability and the occupation probability can be obtained by using $f_v$ in Eqs. \ref{transition prob matrix} and \ref{occprob1} respectively. From these, the load and flux distributions as well as their extreme event probabilities can be computed as was done earlier for degree biased walks. 

\begin{figure} 
\includegraphics{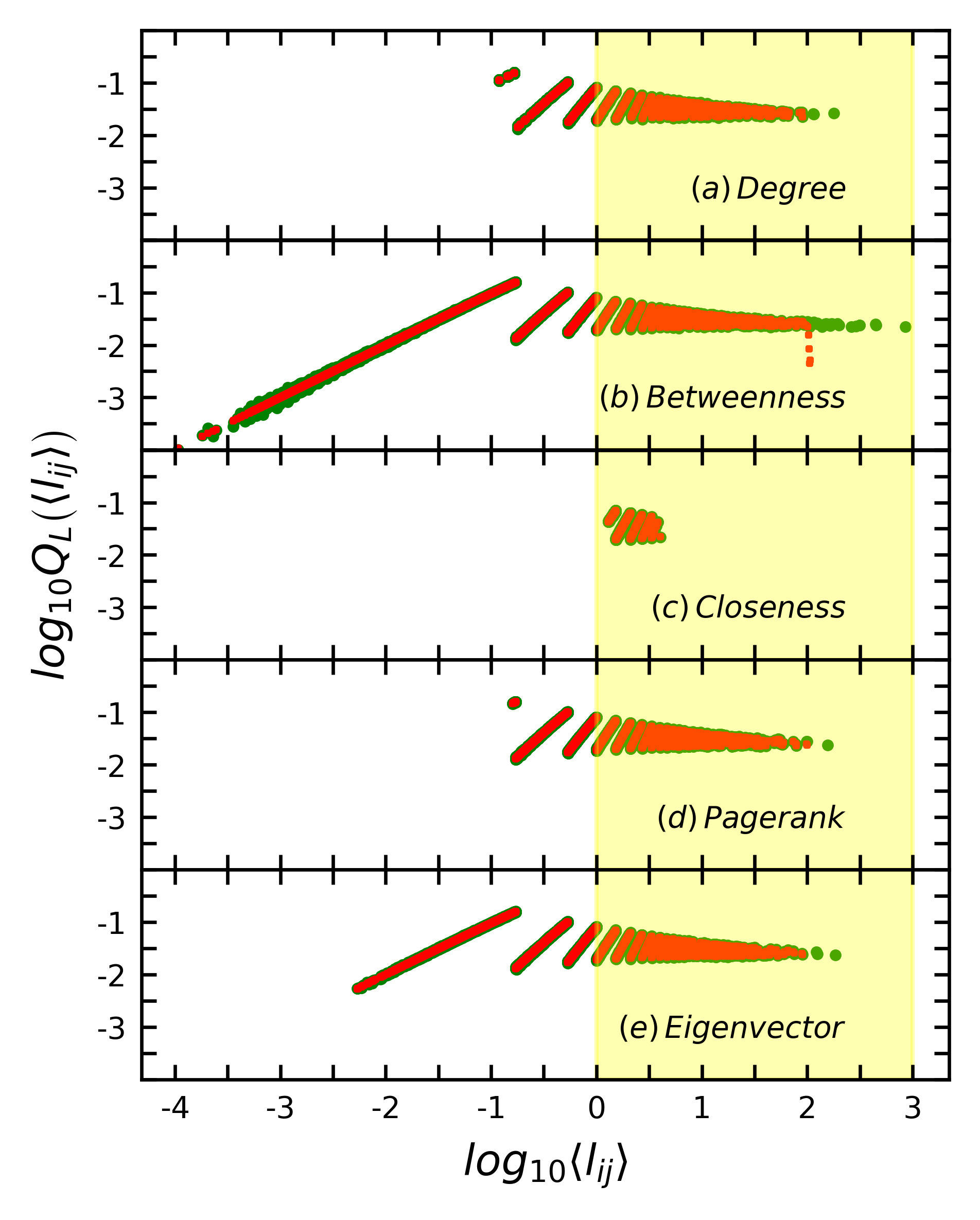}
\caption{Log-log plot of probabilities of edge EE for load against mean edge load using various biasing functions from (a) Degree, (b) Betweenness, (c) Closeness, (d) PageRank, and (e) Eigenvector centralities. The red squares represent analytical results and the green circles indicate numerical simulations. All runs have $m=2$, $\alpha$=1 and were carried out for 75,000 time steps each on a scale-free network with 2000 nodes and 7984 edges with 15968 non-interacting biased random walkers.}
\label{otherbiases}
\end{figure}

In Fig. \ref{otherbiases}(b-e), the load EE probabilities on edge $e_{ij}$ is plotted for the four different biasing functions. For comparison, the degree bias ($k^\alpha$) is shown in Fig. \ref{otherbiases}(a). In all the cases, the same strength parameter $\alpha=1$ and the EE threshold is set to $q_{ij}=\langle l_{ij}\rangle+2\sigma_{ij}$. Even though each of these centrality measures
capture a different facet of the notion of importance for a node, the EE probabilities on edges of the network do not differ significantly irrespective of which biasing function was applied. This can be attributed to correlations found between the centrality measures \cite{oldham2019consistency,valente2008correlated}.

As seen in the case of degree biased random walks, in Fig. \ref{otherbiases} too, a large variability 
in EE probability is observed for edges whose mean load is approximately in the range $\langle l \rangle < 1$, 
as compared to the edges with 
$\langle l \rangle > 1$ (yellow background). Thus, irrespective of the actual biased walk algorithm employed, pronounced
variability in EE probability for some of the edges appears to be a robust feature.
For a given biased dynamics on a network, let 
$\Delta l = \langle l \rangle_{\text max} - \langle l \rangle_{\text min}$
represent difference between the maximum and minimum value of mean loads on its edges. It can be observed
that $\Delta l$ is a function of applied bias. As seen in Figs. \ref{PEEL_2sigma} and \ref{otherbiases}, $\Delta l$ is largest
for betweenness centrality based walk ($ = 8.542 \times 10\textsuperscript{2}$), and smallest for closeness centrality based walk ($ = 2.726 \times 10\textsuperscript{0}$). For unbiased
random walks, $\Delta l = 0$.
This doesn't preclude the possibility that the same edge under different biases can have different EE probabilities. Further, since the EE thresholds are defined based on these measures, it is evident that what might be extreme in one setting 
would theoretically not be so in another. This implies that the choice of bias and its strength are crucial factors
in determining extreme event occurrence rates.

\section{Correlations between Extreme Events} \label{corr_EE}
\begin{figure}
\includegraphics{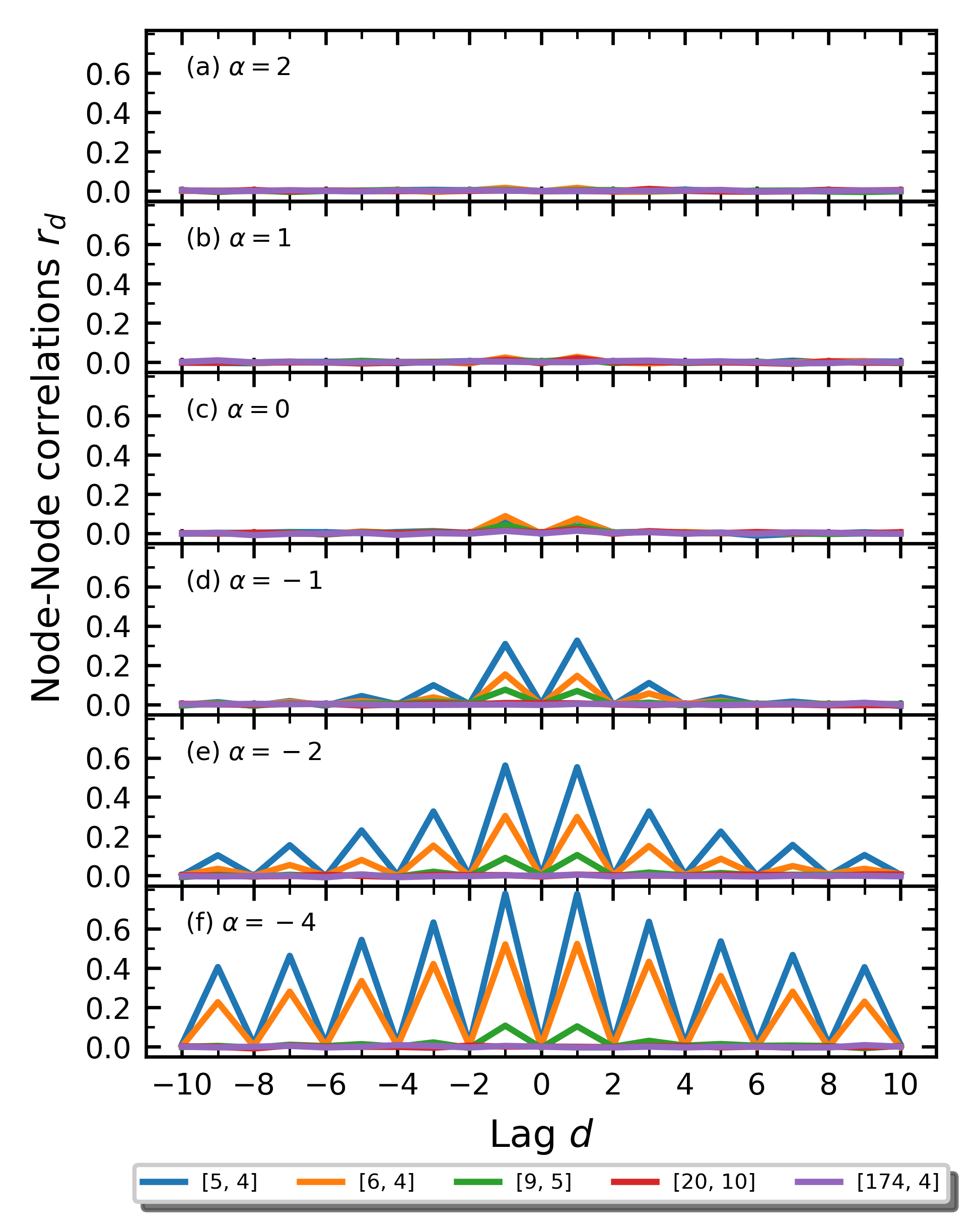}
\caption{Node-Node correlations of EE time series for five representative node pairs labelled by their degrees [$k_i,k_j$] in the legend. Subplots compare this over different strengths of the degree bias (a) $\alpha=2$, (b) $\alpha=1$, (c) $\alpha=1$, (d) $\alpha=-1$, (e) $\alpha=-2$ and (f) $\alpha=-4$ against lag $d$. }
\label{nodenode}
\end{figure}

As reported in \cite{kumar2020extreme}, the unbiased random walk process on networks is uncorrelated, the extreme events on nodes are almost temporally uncorrelated with that on the edges. Yet, the extreme events arising from the 
{\it biased} random walks can display significant time-lagged correlations, {\it i.e.}, extreme events
occurring on nodes can be correlated with the extreme events on the neighbouring nodes or edges after 
a time lag $d$. Let us denote by $\mathcal{N}(i)$ the set of nodes that are connected to node $i$.
In this case, two 
possible scenarios exist, ({\it i}) node-node correlation: EE occurs at time $t$ on a node $i$, and occurs 
at time $t+d$ on node $j$ such that $j \in \mathcal{N}(i)$, ({\it ii}) edge-node correlation: EE occurs at 
time $t$ on an edge $e_{ij}$, and occurs at time $t+d$ on $i$-th or $j$-th node connected
to this edge. To quantify these correlations, we coarse-grain
the time series of walkers as follows. Let $w(t),\, t=1,2,3,\dots,75000$ represent the time varying walker flux through an edge or node.
This series is modified as follows. If an EE occurs at time $t=\tau$, then $w(\tau)=1$, else $w(\tau)=0$. This coarse-grained time series is used to calculate the (normalized) time lagged cross correlation (TLCC) function \cite{gubner2006probability} between any two time series $x(t)$ and $y(t)$. 
It is defined as
\begin{equation}
r_{d}=\frac{\sum_{t=1}^{T-d} (x(t)-\langle x\rangle) ~ (y(t+d)-\langle y\rangle)}{\sqrt{\sum_{t=1}^{T-d}(x(t)-\langle x\rangle)^{2}} ~ \sqrt{\sum_{t=1}^{T-d}(y(t+d)-\langle y\rangle)^{2}} } .
\end{equation}
Here $r_d \in [-1,1]$ and indicates if the two time series are correlated or anti-correlated at time lag $d$. We calculate node-node and edge-node correlation $r_d$ for the EE time series $w(t)$ for all pairs of neighbors in the network with DBRW dynamics for lags $d \in [-10,10]$. The results are shown in Fig. \ref{nodenode} are for five
node pairs. The five node pairs are representative of the different behaviour found, they are labelled with degrees of the nodes the edges connect. Each colour represents the same node pair across all subplots. As is evident in 
Fig. \ref{nodenode}(c), the node-node correlations for unbiased walkers ($\alpha=0$) are barely significant for any time lag $d$. On the other hand, as the bias strength increases, {\it i.e}, $|\alpha| > 0$,
strong correlations emerge among connected pairs of nodes.

From Fig. \ref{nodenode}, we see that the node pairs display strong correlations for lags  $d=-1$ and $d=1$. 
This means that if an EE occurred in a node its neighboring node is likely to have experienced one in the previous time step and/or experience one in the next step. The likelihood of this happening increases as $\alpha$ gets more negative. Furthermore, there exist node pairs (such as the one labelled [$k_i,k_j$]=[5,4] in Fig. \ref{nodenode}(d-f)) whose correlations do not decay rapidly with lag $d$. This means that EE among neighbors can remain correlated for long times. This might be useful for predicting such events. As the bias gets more negative the fraction of pairs with a nontrivial amount of correlation ($\>0.4$ by convention) increases. This fraction depends on degree distribution of the network and for this scale-free network it is $<10\%$ until $\alpha=-2$.  The strength of this lagged correlation depends on
the product of the degrees of the nodes the edge $e_{ij}$ connects, $k_ik_j$. If $k_ik_j$ is large, the
lagged correlation will be small since the EE tend to get "scattered" away in to the many neighbours at the next
time step. If $k_ik_j$ is small, the limited local neighbourhood of poorly connected nodes could result in a fraction of the negatively biased walkers to be dispersed again to the same node.

\begin{figure}
\includegraphics{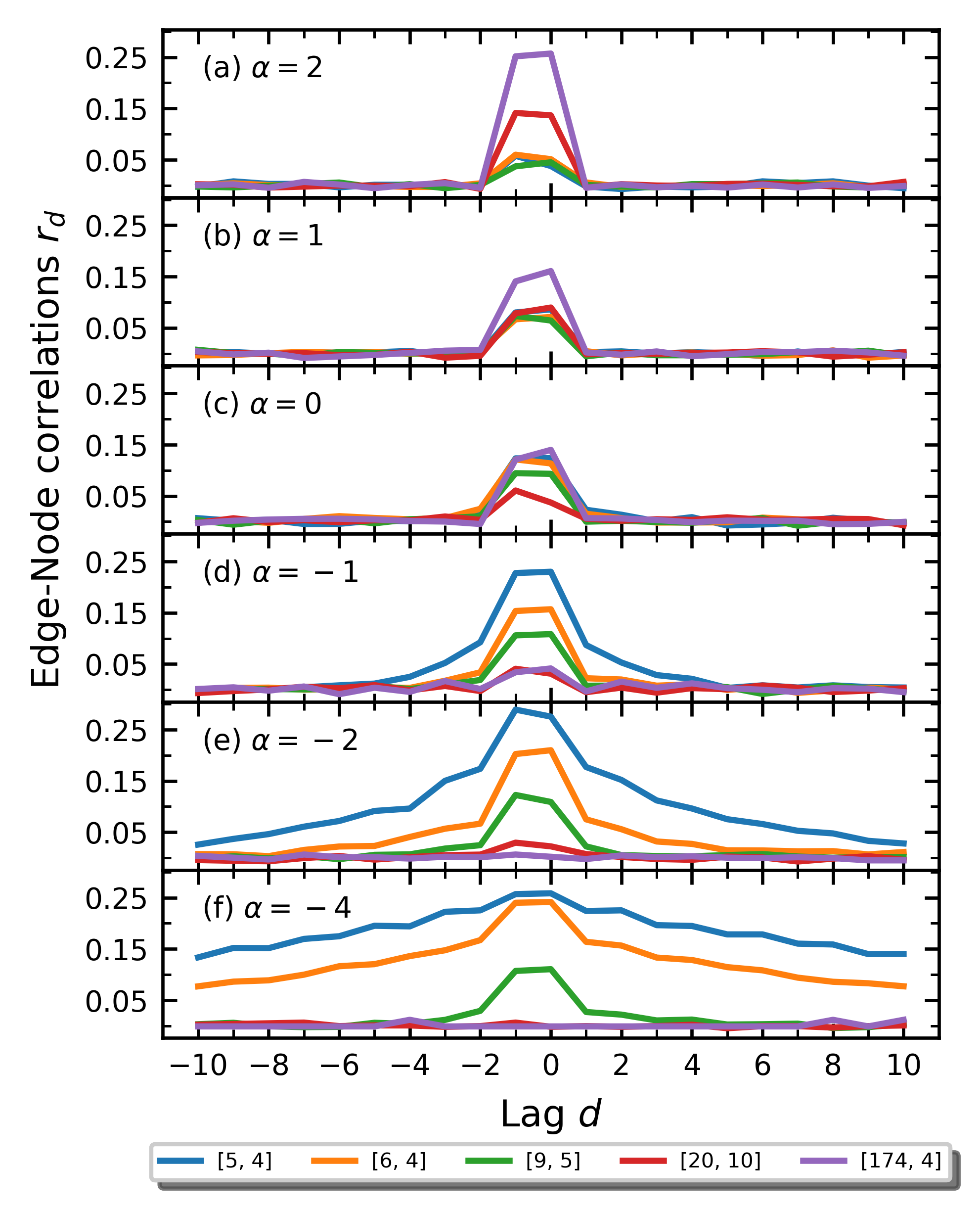}

\caption{Edge-Node correlations between EE time series for five representative edges that were taken and labelled by the degrees of the nodes [$k_i,k_j$] they connected. Subplots compare this over different strengths of the degree bias with bias strength (a) $\alpha=2$, (b) $\alpha=1$, (c) $\alpha=1$, (d) $\alpha=-1$, (e) $\alpha=-2$ and (f) $\alpha=-4$ against lag $d$. The peak values are observed for lag = -1 and 0.}
\label{loadnode}
\end{figure}

Next we investigate the lagged correlation between the EE for load on an edge $e_{ij}$ and the EE 
occurring on one of the nodes it connects to, say, node $i$. In Fig. \ref{loadnode}, lagged correlation $r_d$
is shown for DBRW dynamics for varying bias strength $\alpha$. In general, edge-node correlations are
far weaker than the node-node correlations. For a fixed $\alpha$, the strongest cross correlation 
is seen at time lags of $d=-1$ and $d=1$. It implies that if there is a load EE on an edge then one of 
the nodes connecting it is likely to encounter an EE either one time step before and/or one time step later.
This effect gets more pronounced as $\alpha$ gets more negative. In this case, walkers preferentially
explore small degree nodes which have fewer edges connected to them. Hence, once an extreme event takes place
in one such node, the walkers have no option but to transit through one of its edges creating an extreme event
on the edge as well. Due to this, walkers continually circulate locally among the few small degree nodes 
leading to extremely slow decay of correlations as seen in  Fig. \ref{loadnode}.
As $\alpha$ gets more negative, larger fraction of edges display such slow decay of correlations with lag $d$.

On the other hand, for $\alpha > 0$, the walkers preferentially explore hubs on the
network. If an extreme event takes place in one of the hubs, the walkers are scaterred away through a large
number of edges connected to it. Hence, it might not exceed the threshold required to designate it as an
extreme event. Hence lagged correlations are not pronounced in the case of $\alpha >0$ as observed in 
Fig. \ref{loadnode}(a-b).

\section{Network stability under edge extreme events} \label{stability}
Using the extremes related results obtained above, we are now in a position to discuss the robustness of a network in the context of extreme events faced by edges. To study this, we adopt two broad classes of node or edge deletion strategies: 
({\it i}) targeted node/edge deletion based on decreasing order of EE occurrence probability,
({\it ii}) random edge/node deletion.
To illustrate the effects of these deletion strategies, we monitor the size of the relative 
giant component $S(t)=\frac{N_{\text{max}}(t)}{N}$ in the network as a function of time. It is the 
ratio of nodes in the largest connected component $N_{\text{max}}$ to the total number of nodes $N$ 
{\it initially} in a network. If all the nodes are reachable from any other node, $S=\frac{N_{\text{max}}}{N}=1$. 
As we delete nodes/edges $S(t)$ decreases with time. 

\begin{figure}
\includegraphics{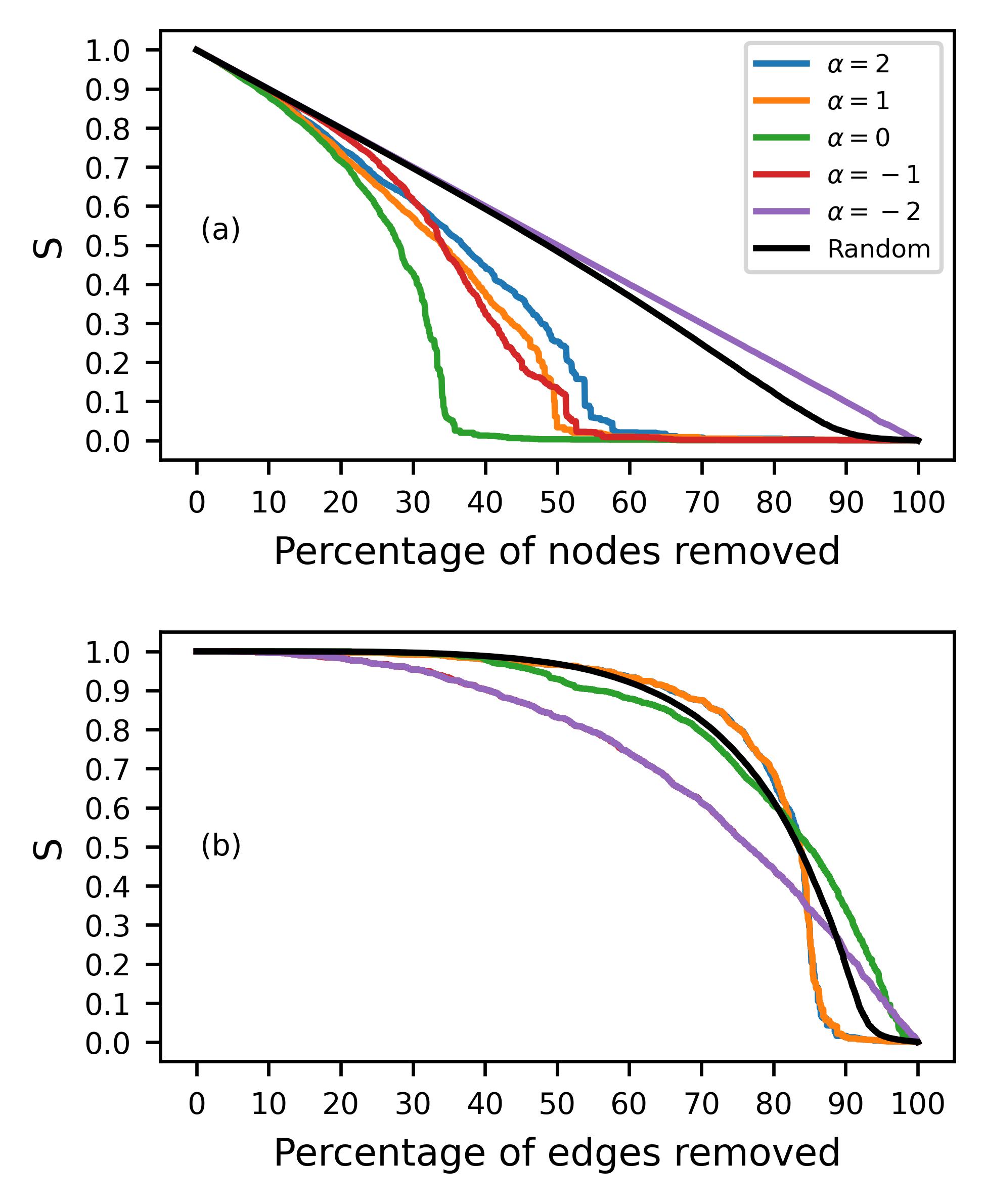}
\caption{Size of the relative giant component tracked as (a) nodes are removed in decreasing order of their node EE probability ${P}_{EE}(q_{i})$ (b) edges are removed in decreasing order of $Q_L \left( \langle l_{ij} \rangle \right)$. In both cases an EE threshold of $m=2$ was used. In (b), curves for $\alpha=-1$ and $-2$ lie on top of each other as do the curves for $\alpha= 1$ and $2$.}
\label{delete}
\end{figure}

In Fig. \ref{delete}(a), the simulation results for a variety of biased random walks are considered
and for each one the targeted and random node deletion strategies are shown. In most cases, even if
30-40\% of the nodes are removed in a targeted manner, the giant component in the network falls to 
less than 50\%. In the case of degree biased random walks, even as 50\% of nodes are targeted and
removed, giant component size becomes vanishingly small. However, if the nodes are randomly
removed, a reasonably sized giant component in the network survives even until about 
80\% of the nodes are deleted. It is well-known that scale-free network is resilient to
random attacks but not to targeted attacks \cite{cohen2001breakdown}. The difference here is that the same viewpoint is reinforced even if the node deletion strategy is based on occurrence of extreme events.

A somewhat surprising behaviour is seen in Fig. \ref{delete}(b), in which edges deletion results
are shown for both targeted and random removal strategies. Unlike the node deletion case,
a significant fraction of the giant component survives well until about 80\% of the edges
are removed. Indeed, until 30\% of edges are deleted, the size of giant component does not
decrease appreciably.  Thus, as far as the survival of network structure is concerned, node and 
edge deletion processes are not equivalent. If $q$\% of nodes 
are deleted, the corresponding giant component size is $S^n_q$. Similarly, if $q$\% of edges are 
removed, let the giant component size be denoted by $S^{e}_q$. Generally, we observe that $S^{e}_q > S^n_q$
for nearly all $q$. Physically, this effect arises because if a node is deleted, typically
many edges becomes dysfunctional for transport of randowm walkers and this tends to quickly 
reduce the size of giant component. On the other hand, if an edge is deleted, most other parts
of the network continues to remain functional for transport. Thus, deletion of a node and an
edge lead to different outcomes for network resilience.
It might also be pointed out that, unlike node deletion strategies, there is virtually
no difference between random and targeted edge deletion for the same reason that scale-free
networks are more resilient to edge deletions than node deletions.

\section{Conclusions}
In summary, we have proposed a general formalism to study extreme events taking place on the edges, rather than on the nodes of a complex network. In this work, we considered the dynamics of biased random walkers on the network, and extreme events on the edges based on the deviation of flux of walkers above a defined threshold. Unlike the case of nodes, in the edge $e_{ij}$ of the network, walkers can go from node $i$ to node $j$ and vice versa. Hence, it was necessary to distinguish between two distinct dynamical quantities -- the load, which is the total number of walkers traversing the edge. Secondly, the flux denoting the difference between the number of walkers going $i$ to $j$ or in the reverse direction. Thus, enabling the consideration of extreme events using both of these dynamical observables. In this work,
we have applied the proposed formalism to study the occurrence of extreme events in load and flux on the edges. It must be pointed out that extreme events in this work represent one possible outcome due to intrinsic fluctuations in the number of biased random walkers, and is not triggered by external sources.

We obtained analytical and numerical estimates for the occurrence probabilities of extreme events on the edges of a network. The occurrence probability of extreme events on edges depends on the local network structure around the edge of interest. Within the framework of biased random walkers, it is shown that the edges can be characterised by the mean loads (or mean fluxes). 
Then, $e_{ij} = e_{ij}(\langle l \rangle)$, where $\langle l \rangle$ denotes mean load. For the edges with low mean loads ($\langle l \rangle < 1$), it is shown that the variability of the extreme event probability is higher compared to edges with higher mean loads ($\langle l \rangle > 1$). Strongly biased random walks tend to increase the variability for $\langle l \rangle < 1$, and suppress the variability at $\langle l \rangle > 1$. We show these results for varying strengths of the degree biased random walk on scale-free networks. We also show similar trends are realised for other biases -- Betweenness, Closeness, Eigenvector, and PageRank centrality based biases. In addition, these are also demonstrated for a real-life planar network.

Further, we studied how the extreme events on nodes and edges are related. Earlier, it was shown that these two classes of extreme events are uncorrelated for unbiased random walkers on networks. However, while using biased random walkers a non-trivial correlation is found to exist. This implies that
an extreme event on a node tends to influence the occurrence or non-occurrence of an extreme event on an edge connected to it. We quantify such correlations for various strengths of the degree bias.

Finally, we also studied the vulnerability of the network as a whole to the occurrence of extreme events
on the edges. We removing nodes/edges in decreasing order of the probability of their extreme event occurrence. For instance, the edge with the highest probability will be removed first, and then the node with next highest probability will be removed and so on. After each removal, the size of
the largest connected component of the network is computed. It is shown that the largest connected component stays intact almost until 50\% of the edges are removed due to extreme events. This is in contrast to the case of node removal based on similar criteria outlined above.
In which case, the size of the giant component falls almost by 50\% when half of the nodes are
removed. Thus, networks can be expected to maintain their resilience much more in the face of
extreme events on the edges, than if it were to happen on the nodes. 

It would be interesting to study extreme events on real-life network with measured loads and fluxes and compare them to patterns observed in synthetic networks. An absorbing exploration would be to map community structure \cite{zlatic2010topologically} but when supplied with the EE data. Keeping track of how EE cascades through the network using higher order versions of the correlations observed among the nodes and edges is another avenue of research. Using measures that quantify the local neighbourhood of nodes involved in node-node and edge-node correlations instead of $k_ikj$ can provide better insight about strong correlations observed. Developing a more accommodating definition of EE can lead to modeling more than congestion-type systems. Using time evolving properties of nodes, edges (and if possible even the walkers) can lead to interesting generalisations of the theory.
Trying out other biasing functions, especially non-structural ones whose values are bestowed externally could lead to interesting trends. One possible extension to this framework is to get equations analytically for time-dependent biasing functions as well. In general, dynamics on the edges have long been neglected and these results, we hope, might lead to more work on extreme events on the edges of networks. 

\begin{acknowledgments}
GG thanks Aanjaneya Kumar for many discussions during this work. GG acknowledges support from IISER Pune and INSPIRE Scholarship for Higher Education (SHE). MSS acknowledges the support from MATRICS grant of SERB, Govt of India.
\end{acknowledgments}

\appendix
\section{Distribution of load over an edge\label{AppB}}

We have $W$ total walkers. The total number of walkers on node $i$
and node $j$ is $n$. The number of walkers that traverse edge
$e_{ij}$ is $l_{ij}$ (load). There are three kinds of events that can happen
in a network from the point of view of an edge $e_{ij}$:
\begin{enumerate}
\item The walker could be neither on node $i$ nor on node $j$. There are $W-n$ such walkers with probability $=1-p_{i} - p_{j} \equiv E_{1}$.
\item The walker could be on node $i$ and not traversing to node $j$; or the walker could be on node $j$ and does not travel to node $i$. There are $n-l_{ij}$ such walkers with probability $=p_{i}(1-\pi_{ij})+p_{j}(1-\pi_{ji}) \equiv E_{2}$.  
\item The walker could be on node $i$ or node $j$ and traverses to node $j$ or node $i$ respectively. There are $l_{ij}$ such walkers with probability $=p_{i}\pi_{ij}+p_{j}\pi_{ji}=2p_{i}\pi_{ij} \equiv E_{3}.$\label{enu:Finally-when-the} 
\end{enumerate}

The detailed balance condition holds for ergodic markov chains {\it i.e.}, $p_{i}\pi_{ij}=p_{j}\pi_{ji}$. The probability that there are $W-n$ events of case 1, $n-l_{ij}$
events of case 2, and $l_{ij}$ events of case 3 happening for an edge $e_{ij}$ is
\begin{equation}
\frac{W!}{(W-n)!(n-l_{ij})!(l_{ij})!}[E_{1}]^{W-n}[E_{2}]^{n-l_{ij}}[E_{3}]^{l_{ij}}\label{b},
\end{equation}
where the $\frac{W!}{(W-n)!(n-l)!(l)!}$ factor accounts for repeated events since the order in which these three independent events occur does not matter. Now, this is just the joint distribution of three events occurring. Since we are not interested in the distribution of walkers among cases 1 and 2. The probability that an edge has load $=l_{ij}$ is simply the marginal probability of Eq. \ref{b} for case 3 to occur {\it i.e.}, the probability that there are $l_{ij}$ walkers that hop from node $i$ to node $j$ or vice versa. This is given by 
\begin{eqnarray}
P_L(l_{ij})=\sum_{n=l_{ij}}^{W}\frac{W!}{(W-n)!(n-l_{ij})!(l_{ij})!}\nonumber \\
\times [E_{1}]^{W-n}[E_{2}]^{n-l_{ij}}[E_{3}]^{l_{ij}}
\end{eqnarray}
After some simple manipulations, we find ourselves with the load distribution from Eq. \ref{prdistload}
\begin{eqnarray}
P_L(l_{ij})=\binom{W}{l_{ij}}(E_{3})^{l_{ij}}(1-E_{3})^{W-l_{ij}}, \nonumber\\
=\binom{W}{l_{ij}}(2p_{i}\pi_{ij})^{l_{ij}}(1-2p_{i}\pi_{ij})^{W-l_{ij}}.
\label{eq:finalee}
\end{eqnarray}


\bibliography{Cits}

\end{document}